\documentclass{amsart}
\usepackage{amsmath}
\usepackage{amssymb}
\catcode`\@=11

\@addtoreset{equation}{section} \catcode`\@=12 
 \DeclareMathOperator{\Z}{\mathbb{Z}}
 \newtheorem*{theorem}{Theorem}

\begin{document}

\title[Hierarchy of quantum models]{Hierarchy of quantum explicitly solvable and integrable
models}
\author{A.~K.~Pogrebkov}
\address{Steklov Mathematical Institute, Moscow, Russia}
\email{pogreb@mi.ras.ru} \curraddr{Steklov Mathematical Institute, Gubkin str., 8, Moscow,
117966, GSP-1, Russia}
\date{\today}
\thanks{This work is supported in part by the Russian Foundation for Basic Research (grants \#
02-01-00484 and 00-15-96046) and by the INTAS (grant \# 99-1782).}

\begin{abstract}
Realizing bosonic field $v(x)$ as current of massless (chiral) fermions we derive hierarchy of
quantum polynomial interactions of the field $v(x)$ that are completely integrable and lead to
linear evolutions for the fermionic field. It is proved that in the classical limit this
hierarchy reduces to the dispersionless KdV hierarchy. Application of our construction to
quantization of generic completely integrable interaction is demonstrated by example of the
mKdV equation.
\end{abstract}
\maketitle

\section{Introduction}
Special quantum fields that first appeared in the literature (see, e.g.~\cite{1}) under the
name ``massless two-dimensional fermionic fields,'' are known for decades to be useful tool of
investigation of completely integrable models in quantum (fermionization
procedure~\cite{2}--\cite{4}) and in classical (symmetry approach to KP hierarchy~\cite{5})
cases. Already in~\cite{2} it was shown that when bosonic field of the quantum version of some
integrable model is considered as a composition of fermions, the most nonlinear parts of the
quantum bosonic Hamiltonian becomes bilinear in terms of these Fermi fields.
In~\cite{6}--\cite{8} the same property was proved for the Nonlinear Schr\"{o}dinger equation
and some integrable models of statistical physics, where fermionic fields naturally appeared
in the so called limit of the infinite interaction, i.e., again as describing the most
nonlinear part of the Hamiltonian. Quantization of the KdV equation is based on analogy of the
Gardner--Zakharov--Faddeev (GZF)~\cite{9} and Magri~\cite{10} Poisson brackets with the
current and Virasoro algebras~\cite{4}, \cite{11}, \cite{12}. In~\cite{4} we proved that
quantization of any of these brackets for the KdV equation by means of fermionization
procedure can be performed on the entire $x$-axis and the Hamiltonian is given as sum of two
terms, bilinear with respect to either fermionic or current operators. We also proved that the
quantum dispersionless KdV equation generates linear evolution equation for the Fermi field.
Thus this equation is explicitly and uniquely solvable for any instant of time (in contrast to
the classical case).

In this article we construct hierarchy of nonlinear interactions for the bosonic quantum field
$v(x)$ that obey the following properties:
\begin{itemize}
\item All equation of this hierarchy are completely integrable in the sense that they have infinite
set of local, polynomial (with respect to $v$ and its derivatives) commuting integrals of
motion.
\item All equations of this hierarchy are explicitly solvable in the following sense. Let $v$ be
realized as current of fermionic field $\psi$. Then all these nonlinear equations for $v$ lead
to linear evolution equations for $\psi$.
\item In the limit $\hbar\to 0$ this hierarchy reduces to the dispersionless KdV hierarchy.
\end{itemize}

The paper is as follows. In Sec.~2 we present some well-known results on the ``two-dimensional
massless'' fermions. In the Sec.~3 the hierarchy is derived and its properties are studied. In
Sec.~4 we demonstrate by means of the modified KdV equation that results of our construction
can be applied to quantization of the generic integrable models. Discussion of the classical
limit of the hierarchy and some concluding remarks are given in the Sec.~5.

\section{Massless two-dimensional fermions}
Here we introduce notations and list some standard properties of the massless Fermi fields
(see, e.g.,~\cite{1}). Let $\mathcal{H}$ denote the fermionic Fock space generated by
operators $\psi(k)$ and $\psi^{*}(k)$, where $*$ means Hermitian conjugation, and that obey
canonical anticommutation relations,
\begin{equation}
\{\psi _{}^{*}(k),\psi (p)\}_{+}^{}=\delta (k-p),\qquad \{\psi
(k),\psi(p)\}_{+}^{}=0.  \label{1}
\end{equation}
Let $\Omega \in \mathcal{H}$ denote vacuum vector and $\psi (k<0)$ and $\psi^{*}(k>0)$ be
annihilation operators,
 \begin{equation}
 \psi (k)\Omega \Bigr|_{k<0}^{}=0,\qquad \psi _{}^{*}(k)\Omega
 \Bigr|_{k>0}^{}=0,  \label{2}
 \end{equation}
whereas $\psi (k>0)$ and $\psi ^{*}(k<0)$ are creation operators. Fermionic
field is the Fourier transform,
 \begin{equation}
 \psi (x)=\frac{1}{\sqrt{2\pi }}\int dk\,e_{}^{ikx}\psi (k),  \label{3}
 \end{equation}
and obeys relations
 \begin{align}
 &\{\psi _{}^{*}(x),\psi (y)\}_{+}^{}=\delta (x-y),\qquad \{\psi(x),\psi(y)\}_{+}^{}=0,
 \label{4} \\
 &(\Omega ,\psi (x)\psi _{}^{*}(y)\Omega)=(\Omega ,\psi_{}^{*}(x)\psi(y)\Omega )=
 \frac{-i\varepsilon_{}^2}{x-y-i0},  \label{5}
\intertext{where we denoted $\varepsilon=(\sqrt{2\pi})^{-1}$. This notation is convenient as
in order to restore the Plank constant $\hbar$ we need not only to substitute all commutators
and anticommutators $[\cdot,\cdot]\to[\cdot,\cdot]\hbar^{-1} $, but also put}
 &\varepsilon=\sqrt{\frac{\hbar}{2\pi}}.  \label{9}
 \end{align}
The current of the massless two-dimensional fermionic field is given by the bilinear
combination
 \begin{equation}
 v(x)=\varepsilon^{-1}_{}\colon\psi _{}^{*}\psi \colon(x), \label{6}
 \end{equation}
where the sign $\colon\ldots \colon$ denotes the Wick ordering with respect to the fermionic
creation--annihilation operators, for example,
 $\colon\psi^*(x)\psi(y)\colon=\psi^*(x)\psi(y)-(\Omega,\psi^{*}(x)\psi (x)\Omega)$ and
 $\colon\psi^{*}\psi \colon(x)=\lim_{y\rightarrow x}\colon\psi^{*}(x)\psi (y)\colon$, etc.
Current is a self-adjoint operator-valued distribution in the space $\mathcal{H}$ obeying the
following commutation relations:
 \begin{align}
 &\lbrack \psi (x),v(y)]=\varepsilon^{-1}_{}\delta (x-y)\psi (x),  \label{7} \\
 &\lbrack v(x),v(y)] =i\delta'(x-y).  \label{8}
 \end{align}
The charge of the fermionic field, $\Lambda =\int dx\,v(x)$, is self-adjoint operator with
spectrum $\varepsilon^{-1}\Z$.

Commutation relation~(\ref{8}) suggests interpretation of $v(x)$ as bosonic field that obeys
quantized version of the GZF bracket (\cite{9}, see also~(\ref{bra}) below). In what follows,
we use the decomposition
 \begin{equation}
 v(x)=v_{}^{+}(x)+v_{}^{-}(x)\label{12}
 \end{equation}
 of this field, where positive and negative parts equal
 \begin{equation}
 v_{}^{\pm }(x)=\frac{\pm 1}{2\pi i}\int \frac{dy\,v(y)}{y-x\mp i0} \label{13}
 \end{equation}
and admit analytic continuation in the upper and bottom half-planes of variable $x$,
correspondingly. They are mutually conjugate and
 \begin{equation}
 v^{-}(x)\Omega =0.  \label{14}
 \end{equation}
Let
\begin{equation}
v(k)=\int dx\,e_{}^{-ikx}v(x), \label{140}
\end{equation}
so that $v^\pm(x)=(2\pi)^{-1}\int dk\,e_{}^{ikx}\theta(\pm k)v(k)$, where $\theta(k)$ is step
function. Then
\begin{equation}
v_{}^{*}(k)=v(-k),\qquad v(k)\Omega \bigl|_{k<0}^{}=0. \label{140:1}
\end{equation}
Thus  $v(k>0)$ and $v(k<0)$  are bosonic creation and annihilation operators, correspondingly,
that are bilinear with respect to fermionic ones. One can introduce the bosonic Wick ordering
for the products of currents, which we denote by the symbol $\vdots \ldots \vdots$, that means
that all positive components of the currents are placed to the left from the negative
components, for instance,
\begin{equation}
\vdots v(x)v(y)\vdots=v_{}^{+}(x)v_{}^{+}(y)+v_{}^{+}(x)v_{}^{-}(y)+v_{}^{+}(y)v_{}^{-}(x)+
v_{}^{-}(x)v_{}^{-}(y)  \label{17}
\end{equation}
and again $\vdots v^{2}\vdots (x)=\lim_{y\rightarrow x}\vdots v(x)v(y)\vdots $. We can also
use equality
\begin{equation}
\vdots v(x)v(y)\vdots =v(x)v(y)-(\Omega ,v(x)v(y)\Omega ),  \label{15}
\end{equation}
where
\begin{equation}
(\Omega ,v(x)v(y)\Omega )=\left(\frac{i\varepsilon}{x-y-i0}\right)_{}^{2}. \label{16}
\end{equation}

Fermionization procedure is essentially based on the relation between these two normal
orderings. The bosonic ordering $\vdots \ldots \vdots $ can be extended for expressions that
include fermionic field:
 \begin{equation}
 \vdots v(x)\psi (y)\vdots =v_{}^{+}(x)\psi (y)+\psi (y)v_{}^{-}(x). \label{16:1}
 \end{equation}
Then by~(\ref{7}) and~(\ref{13}),
 \begin{equation}
 \vdots v(x)\psi (y)\vdots =\colon v(x)\psi (y)\colon+i\varepsilon\frac{\psi(x)-\psi (y)}{x-y},
 \label{16:2}
 \end{equation}
so that this expression as well as its derivatives w.r.t.\ $x$ and $y$ are well-defined in the
limit $y\rightarrow x$. In this limit one uses the obvious fact that under the sign of the
fermionic normal product any expression of the kind
$\colon\ldots\psi(x)\ldots\psi(x)\ldots\colon$ equals to zero. In particular, we get relation
 \begin{equation}
 \vdots v\psi \vdots (x)=i\varepsilon\psi _{x}^{}(x),  \label{16:3}
 \end{equation}
that results in the bosonization of fermions~\cite{2}--\cite{3}. More exactly, one can
integrate this equality and write (at least formally) that
 \begin{equation}
 \psi (x)=\vdots e_{}^{-i\varepsilon^{-1}\int^{x}v(x)dx}\vdots\equiv
 e_{}^{-i\varepsilon^{-1}\int^{x}v^{+}(x)dx}e_{}^{-i\varepsilon^{-1}\int^{x}v^{-}(x)dx},
 \label{16:4}
 \end{equation}
where in the second equality definition of the bosonic normal product was used for the
exponent. Relation~(\ref{16:4}) needs special infrared regularization of the primitive of the
current, $\int^{x}v(x)dx$, and its positive and negative components. This procedure can be
performed, say, like in~\cite{3}, and it leads to a special constant operator conjugated to
the charge $\Lambda$, that must be included in the r.h.s.\ of~(\ref{16:4}).

\section{Hierarchy of explicitly solvable models}

Problems of interpretation of Eq.~(\ref{16:4}) do not appear if we deal with bilinear
combinations of fermionic fields of the type~(\ref{6}). In this case neither infrared
regularization, nor the above mentioned auxiliary operator are needed and product of Fermi
fields is given directly in terms of the current. An analog of such relation is known in the
literature on the symmetries of the KP and KdV hierarchies (see~\cite{5}) in the sense of
formal series.

\begin{theorem} For any real $x$ and $y$ the identity
 \begin{equation}
 \varepsilon^{-1}_{}\colon\psi _{}^{*}(x+y)\psi(x-y)\colon=\varepsilon\frac{\vdots\displaystyle
 \exp\left(i\varepsilon^{-1}\int_{x-y}^{x+y}dx'\,v(x')\right) \vdots -1}{2iy} \label{21}
 \end{equation}
holds in the sense of operator-valued distributions of spatial variable $x$ that smoothly
depend on parameter $y$.
\end{theorem}

{\sl Proof.\/} Thanks to (\ref{16:1}) and (\ref{16:3}) we have that
 \begin{align*}
 & -i\varepsilon\frac{\partial }{\partial y}\psi _{}^{*}(x+y)\psi(x-y)=\\
 &\quad=v_{}^{+}(x+y)\psi_{}^{*}(x+y)\psi(x-y)+\psi_{}^{*}(x+y)\psi (x-y)v_{}^{-}(x-y)+\\
 &\qquad+\psi_{}^{*}(x+y)(v_{}^{-}(x+y)+v_{}^{+}(x-y)) \psi (x-y)\equiv \\
 & \quad\equiv (v_{}^{+}(x+y)+v_{}^{+}(x-y))\psi _{}^{*}(x+y)\psi (x-y)+\\
 &\qquad+\psi_{}^{*}(x+y)\psi (x-y)(v_{}^{-}(x+y)+v_{}^{-}(x-y))+ \\
 & \qquad+[\psi _{}^{*}(x+y),v_{}^{+}(x-y)]\psi (x-y)+\psi _{}^{*}(x+y)[v_{}^{-}(x+y),\psi (x-y)]
 \end{align*}
Taking~(\ref{7}) and~(\ref{13}) into account we get
 \begin{align}
 &-i\varepsilon\frac{\partial }{\partial y}\psi _{}^{*}(x+y)\psi (x-y)=\label{18:1}\\
 &\quad=\left( v_{}^{+}(x+y)+v_{}^{+}(x-y)+\frac{i\varepsilon}{2(y-i0)}\right) \psi
 _{}^{*}(x+y)\psi (x-y)+ \nonumber\\
 &\qquad +\psi _{}^{*}(x+y)\psi (x-y)\left( v_{}^{-}(x+y)+v_{}^{-}(x-y)+\frac{i\varepsilon}
 {2(y-i0)}\right)\nonumber
 \end{align}
Let
 \begin{equation}
 F(x,y)=\varepsilon^{-1}_{}\colon\psi _{}^{*}(x+y)\psi (x-y)\colon,  \label{18}
 \end{equation}
i.e., $F$ denotes the l.h.s.\ of~(\ref{21}). Then by definition of the fermionic Wick ordering
we get $\psi _{}^{*}(x+y)\psi (x-y)=\varepsilon F(x,y)-i\varepsilon^{2}(2y-i0)^{-1}$. We
substitute $\psi^*\psi$ in~(\ref{18:1}) by means of this equality and take into account that
powers of distributions $(y-i0)^{-1}$ are well defined. Then we get the following differential
equation for $F(x,y)$:
 \[
 \frac{\partial F(x,y)}{\partial y}=\frac{i}{\varepsilon}\vdots \bigl( v(x+y)+v(x-y)\bigr)
 F(x,y)\vdots +\frac{v(x+y)+v(x-y)-2F(x,y)}{2(y-i0)}.
 \]
At the same time by~(\ref{6}) and~(\ref{18}) at $y=0$
 \begin{equation}
 F(x,0)=v(x),  \label{18:3}
 \end{equation}
so the $i0$ term in the denominator can be omitted and we get
 \begin{equation}
 \frac{\partial F(x,y)}{\partial y}=\frac{i}{\varepsilon}\vdots \bigl( v(x+y)+v(x-y)\bigr)
 F(x,y)\vdots +\frac{v(x+y)+v(x-y)-2F(x,y)}{2y}.
 \label{18:2}
 \end{equation}
It is easy to check directly that the r.h.s.\ of~(\ref{21}) obeys the same differential
equation with respect to $y$ and the same boundary condition~(\ref{18:3}), that
proves~(\ref{21}).

Thanks to~(\ref{18}) it is obvious that $F(x,y)$, indeed, is operator-valued distribution with
respect to $x$ that is smooth, infinitely differentiable function of $y$. Then the same are
properties of this function when it is given by the r.h.s.\ of~(\ref{21}) as
 \begin{equation}
 F(x,y)=\varepsilon\frac{\vdots\displaystyle\exp\left(i\varepsilon^{-1}_{}\int_{x-y}^{x+y}dx'\,v(x')\right)
 \vdots -1}{2iy}\,,
 \label{21:1}
 \end{equation}
that completes proof of the theorem.

Let us introduce
 \begin{equation}
 F_{n}^{}(x)\equiv \left(\frac{\varepsilon\partial_{y}^{}}{2i}\right)_{}^{n}F(x,y)\Bigl|_{y=0}^{}=
 \frac{\varepsilon^{n-1}_{}}{(2i)^n_{}}D_{}^{n}(\colon\psi _{}^{*}\cdot \psi \colon)(x),  \label{22}
 \end{equation}
where in the second equality we used notation for the Hirota derivative~\cite{5}, that in the
generic case of two functions $f(x)$ and $g(x)$ reads as
 \begin{equation}
 D_{}^{n}(f\cdot g)(x)=\lim_{y\rightarrow 0}\frac{\partial _{}^{n}}{\partial
 y_{}^{n}} f(x+y)g(x-y),\quad n=1,2,\ldots .  \label{Hirota}
 \end{equation}
In particular, by~(\ref{18:3}) we get that
 \begin{align}
 &F_{0}^{}(x)=v(x),\label{25}\\
 &F_1^{}(x)=\frac{1}{2i}D(\colon\psi _{}^{*}\cdot \psi\colon)(x),\label{25:1}
 \end{align}
that are current and energy--momentum density of the massless fermi-field, correspondingly.
Thus Eq.~(\ref{21}) gives relation of the Hirota derivatives of the fermionic fields with
polynomials of the current and its derivatives. All $F_n(x)$ by~(\ref{22}) are self-adjoint
operator-valued distributions on the Fock space $\mathcal{H}$ and by~(\ref{21:1}) we get
recursion relations
 \begin{align}
 &F_{2n+1}^{}(x)=\frac{1}{2n+2}\sum_{m=0}^{n}\frac{(-i\varepsilon/2)_{}^{2m}(2n+1)!} {(2m)!(2(n-m))!}\,
 \vdots v_{}^{(2m)}(x)F_{2(n-m)}^{}(x)\vdots\,,  \nonumber \\
 &\qquad\qquad\qquad\qquad\quad n=0,1,2,\ldots ,  \label{23} \\
 \intertext{and}
 &F_{2n}^{}(x) =\frac{1}{2n+1}\sum_{m=0}^{n-1}\frac{(-i\varepsilon/2)_{}^{2m}(2n)!}
 {(2m)!(2(n-m)-1)!}\,\vdots v_{}^{(2m)}(x)F_{2(n-m)-1}^{}(x)\vdots+{}  \nonumber \\
 &\qquad\qquad\qquad\qquad +\left(\frac{\varepsilon}{2i}\right)_{}^n
 \frac{v_{}^{(2n)}(x)}{2n+1},\quad n=1,2,3,\ldots,\label{24}
 \end{align}
where $F_0$ is given in~(\ref{25}). The lowest simplest examples are as follows:
 \begin{align}
 &F_1^{}(x)=\frac{1}{2}\vdots v_{}^{2}\vdots(x),  \label{26} \\
 &F_2^{}(x)=\frac{1}{3}\vdots v_{}^{3}\vdots(x) -\frac{\varepsilon^2_{}v^{}_{xx}(x)}{12},
 \label{27} \\
 &F_3^{}(x)=\frac{1}{4}\vdots v_{}^4\vdots (x)-\frac{\varepsilon^2_{}}{4}\vdots v(x)v^{}_{xx}(x)\vdots, \label{28} \\
 &F_4^{}(x)=\frac{1}{5}\vdots v^{5}_{}\vdots (x)-\frac{\varepsilon^2_{}}{2} \vdots
 v_{}^{2}(x)v^{}_{xx}(x)\vdots + \frac {\varepsilon^4_{}v^{}_{xxxx}(x)}{80}.  \label{29}
 \end{align}

By definition~(\ref{18}) operator $F(x,y)$ obeys commutation relation
 \begin{align}
 [F(x,y),F(x',y')]=& -\varepsilon^{-1}_{}\delta (x-x'+y+y')F(x+y',y+y')+{}  \label{33}\\
 & +\varepsilon^{-1}_{}\delta (x-x'-y-y')F(x'+y,y+y')+{} \nonumber\\
 & +i\frac{\delta (x-x'+y+y')-\delta (x-x'-y-y')}{y+y'},\nonumber
 \end{align}
that generates corresponding commutation relations for $F_{m}$ (closely related with a
representation of the $\mathfrak{gl}_\infty$-algebra). Only the lowest terms, $F_0$ and $F_1$,
form closed subalgebras:
 \begin{align}
 [F_{0}^{}(x),F_{0}^{}(x')] &=i\delta'(x-x'),  \label{34} \\
 [F_{0}^{}(x),F_{1}^{}(x')] &=i\delta'(x-x')F_{0}^{}(x'),  \label{35} \\
 [F_{1}^{}(x),F_{1}^{}(x')] &=i\{F_{1}^{}(x)+F_{1}^{}(x')\}\delta'(x-x')
 -\frac{i\varepsilon^2_{}}{12}\delta'''(x-x'), \label{36}
 \end{align}
while commutators of the type $[F_m,F_n]$ include $F_j$'s till $F_{m+n-1}$.

Operator $F(x,y)$ admits integration with respect to $x$ along the entire axis and result of
integration is well defined operator in the fermionic Fock space $\mathcal{H}$. Indeed,
by~(\ref{3}) and~(\ref{18})
 \begin{equation}
 \int dx\,F(x,y)=\frac{1}{\varepsilon}\int\limits_{0}^{\infty}dk\,\left(e_{}^{2iky}\psi_{}^{*}(-k)
 \psi(-k)-e_{}^{-2iky}\psi(k)\psi _{}^{*}(k)\right) , \label{38}
 \end{equation}
where expression in the r.h.s.\ is normally ordered and has creation$\times$annihilation form,
so that thanks to~(\ref{2})
 \begin{equation}
 \int dx\,F(x,y)\,\Omega =0  \label{38:1}
 \end{equation}
for any $y$. From here we derive that all operators
 \begin{equation}
 H^{}_n\equiv\int dx\,F_{n}^{}(x)=\frac{1}{\varepsilon}\int\limits_{0}^{\infty }dk\,
 (\varepsilon k)_{}^{n}\bigl(\psi_{}^{*}(-k)\psi(-k)-(-1)_{}^{n}\psi(k)\psi _{}^{*}(k)\bigr)
 \label{39}
 \end{equation}
are well defined and self-adjoint. For odd $n$ they are positively defined. At the same time
by~(\ref{33}) we get
 \begin{equation}
 \left[ \int dx\,F(x,y),\int dx'\,F(x',y')\right] =0\label{40}
 \end{equation}
for any $y$ and $y'$. This means in particular that all
 \begin{equation}
 [ H^{}_m,H^{}_n] =0,\quad m,n=0,1,\ldots \label{40:1}
 \end{equation}
In other words, these operators define commuting flows on the space $\mathcal{H}$ and we can
introduce hierarchy of integrable time evolutions by means of commutation relation
 \begin{equation}
 v^{}_{t_m}(x)=i[ H^{}_m,v(x)],\quad m=0,1,\ldots, \label{40:2}
 \end{equation}
so that by~(\ref {40:1}): $(\partial_{t_m}\partial_{t_n}-\partial_{t_n}\partial_{t_m})v(x)=0$
for any $m$ and $n$ (we do not indicate the time dependence in all cases where it is not
necessary). On the other side, by~(\ref {33})
 \begin{align}
 &\left[ \int dx\,F(x,y),v(x')\right]=\varepsilon^{-1}_{}\bigl[F(x'+y,y)-F(x'-y,y)\bigr]\equiv
 \label{41}\\
 &\qquad\equiv\frac{1}{2iy}\left\{\vdots\exp
 \left(i\varepsilon^{-1}_{}\int\limits_{x'}^{x'+2y}d\xi \,v(\xi )\right) \vdots -\vdots \exp
 \left(i\varepsilon^{-1}_{}\int\limits_{x'-2y}^{x'}d\xi \,v(\xi )\right) \vdots \right\} ,
 \nonumber
 \end{align}
that leads to highly nonlinear (polynomial) dynamic equations for $v(x)$ in all cases with
exception to $t_0$ and $t_1$. Thanks to~(\ref {22}),~(\ref {39}), and~(\ref {40:2}) we have:
 \begin{align}
 &v^{}_{t_0}(x)=0,\label{42}\\
 &v^{}_{t_1}(x)=v^{}_x(x),\label{43}\\
\intertext{and in the generic situation}
 &v^{}_{t_n}(x)=\frac{\partial}{\partial x}\sum_{m=0}^{\left[\frac{n-1}{2}\right]}
 \frac{(i\varepsilon/2)_{}^{2m} n!\,\partial^{2m}_xF^{}_{n-2m-1}(x)}{(n-2m-1)!(2m+1)!},\quad
 n=1,2,\ldots.\label{44}\\
\intertext{The simplest examples are as follows:}
 &v^{}_{t_2}(x)=\partial^{}_x\vdots v^{2}_{}\vdots(x),\label{45}\\
 &v^{}_{t_3}(x)= \partial^{}_x\left(\vdots v_{}^3\vdots (x)-\frac{\varepsilon^2_{}}{2}v^{}_{xx}(x)\right),\label{46}\\
 &v^{}_{t_4}(x)= \partial^{}_x\Bigl(\vdots v_{}^4\vdots (x)-2\varepsilon^2_{}\vdots vv^{}_{xx}\vdots(x)-
 \varepsilon^2_{}\vdots v^{2}_{x}\vdots(x)\Bigr).
 \label{47}
 \end{align}
These polynomial interactions are closely related to the KdV hierarchy: the second evolution
is just dispersionless quantum KdV (cf.~\cite{4}), the third evolution coincide with the
modified KdV equation for some specific value of the interaction constant, and so on. In the
next section we discuss the case of mKdV equation in more detail. Here we emphasize that in
spite of the highly nonlinear form of all these equations in terms of the field $v$, all of
them give linear evolutions for fermions. Indeed, introducing the time dependence of $\psi(x)$
in analogy with~(\ref{40:2}) as $\psi_{t_m}=i[H_m,\psi]$, we get by~(\ref{39})
 \begin{equation}
 \psi_{t_m}^{}(x)=\frac{1}{i\varepsilon}(i\varepsilon\partial^{}_x)_{}^m\psi(x),
 \label{47:1}
 \end{equation}
or by~(\ref{3}) $\psi_{t_m}^{}(k)=(i\varepsilon)^{-1}(-\varepsilon k)_{}^m\psi(k)$. Let now
$\psi(t_m,x)$, $v(t_m,x)$, and $F(t_m,x,y)$ be operators with time evolution given by some
$H_m$ and determined by the condition that at $t_m=0$ they equal to $\psi(x)$, $v(x)$, and
$F(x,y)$, correspondingly. Thanks to~(\ref{38:1}) the definitions of the both normal products
do not depend on time. This means that these operators are related at arbitrary value of $t_m$
by means of the same Eqs.~(\ref{6}), (\ref{18}), (\ref{21:1}), and~(\ref{25}) as at $t_m=0$.
In particular, by~(\ref{18})
 \begin{equation}
 F(t^{}_m,x,y)=\frac{1}{\varepsilon}\colon\psi _{}^{*}(t^{}_m,x+y)\psi (t^{}_m,x-y)\colon\,.  \label{48}
 \end{equation}
Then, thanks to~(\ref{3}), (\ref{18}), and~(\ref{47:1}) we get {\sl explicit} expression for
$F(t_m,x,y)$ in terms of its initial value $F(x,y)$:
\begin{align}
 &F(t^{}_m,x,y)=\frac{2}{(2\pi)_{}^2}\int dx'\int dy'\int dk\int dp\,F(x-x',y-y')\times
 \label{49}\\
 &\qquad\times\exp\bigl(i(k-p)x'+i(k+p)y'+i\varepsilon^{m-1}_{}(k^m-p^m)t_m\bigr).
 \nonumber\\
\intertext{Thanks to~(\ref{22}) and~(\ref{25}) we obtain for $y=0$:}
 &v(t^{}_m,x)=\frac{2}{(2\pi)_{}^{2}}\int dx'\int dy'\int dk\int dp \,
 F(x-x',y')\times \label{50}\\
 &\qquad\times\exp\bigl(i(k-p)x'-i(k+p)y'+i\varepsilon^{m-1}_{}(k^m-p^m)t_m\bigr).
 \nonumber\\
\intertext{Substituting here $F(x,y)$ by means of~(\ref{21:1}) we get solution of the $m$'s
equation of the hierarchy~(\ref{40:2}) in terms of the initial data $v(x)$:}
 &v(t^{}_m,x)=\frac{1}{(2\pi)_{}^{2}}\int dx'\int dy'\int dk\int dp \,\frac{\vdots\displaystyle \exp
 \left(i\varepsilon^{-1}_{}\int_{x-x'-\varepsilon y'}^{x-x'+\varepsilon y'}dx''\,v(x'')\right) \vdots -1}{2iy'}
 \times\label{51}\\
 &\qquad\times\exp\left(ikx'-ipy'+i\frac{t_m}{2^m_{}\varepsilon}[(p+\varepsilon k)^m_{}-(p-\varepsilon k)^m_{})]\right).
 \nonumber
\end{align}
Generalization to the case where time evolution is determined by a linear combination of
Hamiltonians $H_m$ is straightforward.

Thus we see, that all these models are not only completely integrable, but also explicitly
solvable in the fermionic Fock space $\mathcal{H}$. On the other side, taking into account
that thanks to~(\ref{40:1}) and~(\ref{42}) the charge operator $\Lambda=H_0/\sqrt{2\pi}$
commutes with all Hamiltonians and $v(x)$, one can reduce bosonic equations to the zero (or
any other, fixed) charge sector of $\mathcal{H}$, that is exactly the standard bosonic Fock
space. In that case all relations of the type~(\ref{21:1}) and~(\ref{50}) remain valid and
give explicit solution of the hierarchy~(\ref{40:2}) in the bosonic Fock space.

\section{The modified KdV equation}

The modified Korteweg--de Vries (mKdV) equation
 \begin{equation}
 v^{}_{t}=\partial^{}_x\left(gv_{}^{3}-\frac{v^{}_{xx}}{2}\right)  \label{mKdV}
 \end{equation}
for the real function $v(t,x)$ is well known example of the completely integrable differential
equation. If $v(x)$ is a smooth real function that decays rapidly enough when $|x|\rightarrow
\infty $, the Inverse Spectral Transform (IST) method (see~\cite{13, 14} and references
therein) is applicable to Eq.~(\ref{mKdV}). Constant $g$ in this equation is an arbitrary real
parameter and properties of solutions essentially depend on its sign. In particulary, the
soliton solutions exist only if $g<0$.

The mKdV equation is Ha\-mil\-to\-ni\-an system with respect to the GZF bracket~\cite{9},
 \begin{equation}
 \{v(x),v(y)\}=\delta^{\prime}(x-y),  \label{bra}
 \end{equation}
so that Eq.~(\ref{mKdV}) can be written in the form $v_{t}=-\{H,v\}$, where Hamiltonian
 \begin{equation}
 H=\frac{1}{4}\int dx\left( gv_{}^{4}(x)+v_{x}^{2}(x)\right)  \label{H}
 \end{equation}

The direct quantization of the mKdV equation on the whole axis requires some regularization
(e.g., space cut-off) of the Hamiltonian in order to supply it with operator meaning. Any such
regularization is incompatible with the IST already in the classical case: the continuous and
discrete spectra of corresponding linear (Zakharov--Shabat) problem become mixed and the most
interesting, soliton solutions cease to exist.

Here we show that realizing $v(x)$ as in~(\ref{6}), i.e., as a composition of fermionic fields
we can avoid any cut-off procedure in~(\ref{H}), because the Hamiltonian becomes well-defined
in the fermionic Fock space $\mathcal{H}$.

We choose the quantum Hamiltonian to be bosonically ordered expression~(\ref{H}),
 \begin{equation}
 H=\frac{1}{4}\int dx\vdots gv_{}^{4}(x)+ v_{x}^{2}(x)\vdots\, .  \label{1-2}
 \end{equation}
Then, thanks to~(\ref{28}) we get
 \begin{equation}
 H=gH^{}_3+\frac{1-g\varepsilon^{2}_{}}{4} \int dx\vdots
 v_{x}^{2}\vdots (x), \label{1-3}
 \end{equation}
where~(\ref{39}) for $n=3$ was used. Thus, in analogy with the KdV case (see~\cite{4}), the
most singular part of the Hamiltonian~(\ref{1-2}) that was of the fourth order with respect to
bosonic operators is only of the second order with respect to fermions. Taking into account
that by~(\ref{140})
 \begin{equation}
 \int dx\vdots v_{x}^{2}\vdots (x)=2\int\limits_{0}^{\infty }dk\,k_{}^{2}v(k)v(-k) \label{1-5}
 \end{equation}
we get that both terms in~(\ref{1-3}) are bilinear in either fermionic, or bosonic
creation--an\-ni\-hi\-la\-ti\-on operators, they are normally ordered and have a diagonal
form, i.e., they include ``crea\-ti\-on$\times$an\-ni\-hi\-la\-ti\-on'' terms only.
Correspondingly, both these terms are well defined self-adjoint operators in $\mathcal{H}$ and
under our quantization procedure no any regularization of the Hamiltonian is needed. In
particular, by~(\ref {140:1}) and~(\ref{38:1})
 \begin{equation}
 H\Omega =0\label{1-6}
 \end{equation}
and by~(\ref{39}) and~(\ref{1-5}) the Hamiltonian~(\ref{1-3}) is positively defined when
$\varepsilon^{-2}_{}\geq g\geq 0$.

It is clear that time evolution given by the Hamiltonian~(\ref{1-2}),
 \begin{equation}
 v_{t}^{}=i[H,v]\equiv \partial^{}_x\left(g\vdots v_{}^3\vdots -\frac{v_{xx}^{}}{2}
 \right),\label{1-8}
 \end{equation}
is exactly the quantum version of the Eq.~(\ref{mKdV}) normally ordered with respect to the
bosonic operators. Thanks to~(\ref{27}) we can exclude the $v^3$-term and get the quantum
\textbf{bilinear} form of the mKdV equation in terms of the fermionic fields:
 \begin{equation}
 v_{t}^{}(x)=\frac{\partial }{\partial x}\left(3gF^{}_2(x)+\frac{g\varepsilon^{2}_{}-2}{4}
 v_{xx}^{}(x)\right) ,  \label{1-9}
 \end{equation}
that can be considered as a quantum Hirota form of the mKdV equation.

In order to derive time evolution of the fermionic field $\psi$ it is reasonable to rewrite
the second term of~(\ref{1-3}) by means of the fermionic normal ordering. Omitting details we
get by definitions of the both normal orderings and Eqs.~(\ref{6}) and~(\ref{18}) the equality
 \begin{equation}
 \vdots v(x)v(y)\vdots=\colon v(x)v(y)\colon+\varepsilon\frac{\displaystyle F\left(\frac{x+y}{2},
 \frac{x-y}{2}\right)-\displaystyle F\left(\frac{x+y}{2},\frac{y-x}{2}\right)}{i(x-y)},  \label{1-91}
 \end{equation}
that after differentiation gives in the limit $y\to x$
 \begin{equation}
 \vdots v_{x}^{2}\vdots(x)=\colon v_{x}^{2}\colon(x)+\frac{1}{2}\partial^2_xF^{}_1(x)+
 \frac{2}{3\varepsilon^{2}_{}}F^{}_3(x),
 \label{1-92}
 \end{equation}
where~(\ref{22}) was used and where by~(\ref{6})
 $\colon v_{x}^{2}\colon(x)=2\varepsilon^{-2}_{}\colon\psi^*_x\psi^*_{}\psi_x\psi\colon$. Thus we
can write~(\ref{1-3}) as
 \begin{equation}
 H=\frac{5g+\varepsilon^{-2}_{}}{6}H^{}_3+\frac{\varepsilon^{-2}_{}-g}{2}
 \int dx\colon\psi^*_x\psi^*_{}\psi^{}_x\psi\colon(x),
 \label{1-93}
 \end{equation}
and thus time evolution of the fermionic field, $\psi_t=i[H,\psi]$ is given by equation
 \begin{equation}
 \psi _{t}^{}(x)=-\frac{5g\varepsilon^{2}_{}+1}{6}\psi_{xxx}^{}(x)+
 \frac{g\varepsilon^{2}_{}-1}{2i\varepsilon}\colon v_{xx}^{}\psi
 \colon(x),  \label{1-10}
 \end{equation}
that is, of course, nonlinear when $g\neq \varepsilon^{-2}$.

Investigation of the spectrum of the quantum Hamiltonian deserves the separate studying. But
like in the~\cite{4} it can be shown that in the fermionic Fock space $\mathcal{H}$ for $g<0$
there exists one-soliton state, i.e., such state that the average of the field $v$ with
respect to it equals to the classical one-soliton solution at least at zero (or any fixed)
instant of time. This state does not belong to the zero charge sector of $\mathcal{H}$, so it
cannot exist in the standard (bosonic) quantization of the mKdV equation. Again, like
in~\cite{4} it can be shown that existence of this state implies quantization of the soliton
action variable.

\section{Conclusion}

We derived hierarchy of nonlinear integrable and at the same time solvable evolutions of the
bosonic field $v(x)$ realized as composition of the fermionic fields---current. By~(\ref{25})
this means that $F_0(x)$ was chosen to be a dynamical variable. But the closed subalgebra of
commutation relations~(\ref{34})--(\ref{36}) is given also by $F_0(x)$ and $F_1(x)$. Moreover,
the linear combination
 \begin{equation}
 \widetilde{F}(x)=F^{}_{1}(x)+a\partial^{}_xF^{}_{0}(x)\label{36:1}
 \end{equation}
with real constant coefficient $a$ also obeys closed commutation relation,
 \begin{equation}
 [\widetilde{F}(x),\widetilde{F}(x')]=i\{\widetilde{F}(x)+\widetilde{F}(x')\}\delta'(x-x')
 -i\left(a^2_{}+\frac{\varepsilon^2_{}}{12}\right)\delta'''(x-x'), \label{37}
 \end{equation}
as follows from (\ref{34})--(\ref{36}). This means that $\widetilde{F}(x)$ gives another
possible choice of a dynamical variable. In~\cite{4} we proved that the dispersionless KdV in
this case is also solvable, while---in contrast to the above---it was $v(x)$ that evolved
linearly. It is natural to expect that the same property is valid for the entire
hierarchy~(\ref{44}) generated by the quantum version~(\ref{37}) of the Magri bracket.

Coefficients of the r.h.s.\ of the bosonic equations of motion~(\ref{43})--(\ref{47}) are
uniquely (up to a common factor) fixed by recursion relations~(\ref{23})--(\ref{24}). Indeed,
transformation
 \begin{equation}
 v(x)\to av(ax),
 \label{2:0}
 \end{equation}
is the only canonical scaling transformation that is unitary implemented in $\mathcal{H}$.
Here constant $a>0$ in order to preserve definition~(\ref{13}) of positive and negative parts
of $v$. This transformation generates:
 \begin{equation}
 \psi(x)\to \sqrt{a}\psi(ax),\qquad F(x,y)\to aF(ax,ay), \qquad F_n^{}(x)\to
 a^n_{}F^{}_n(ax),
 \label{2:01}
 \end{equation}
that is compatible with~(\ref{25})--(\ref{24}). Thus by~(\ref{39}) $H_n\to a^{n-1}H_n$, and
thanks to~(\ref{44}) transformation~(\ref{2:0}) can be compensated by rescaling  of times:
$t_n\to a^{1-n}t_n$.

Flows given in~(\ref{25})--(\ref{29}) are close to the flows of the KdV hierarchy~\cite{13}:
they are polynomial with respect to $v(x)$ and its derivatives and have the same leading
terms. On the other side, the lowest nontrivial example~(\ref{27}) shows that some essential
terms that are involved in the KdV case are absent in~(\ref{44}). In fact, as it was natural
to expect by~\cite{4}, Eq.~(\ref{27}) is the dispersionless KdV equation: the term
$v_{xxx}(x)$ is absent. The higher equations, like~(\ref{28}), (\ref{29}), and so on already
include terms with derivatives, so these equations are not the dispersionless ones. On the
other side, coefficients of all such terms of all commuting flows introduced in Sec.~3 are
proportional to powers of $\varepsilon^2$, i.e., of $\hbar$ by~(\ref{9}). Thanks to~(\ref{25})
and~(\ref{23}), (\ref{24}) it is easy to see that in the limit $\hbar\to 0$
 \begin{equation}
 F^{}_m(x)\to \frac{v^{m+1}_{}(x)}{m+1}, \label{2:1}
 \end{equation}
so that by~(\ref{44}) we get in the classical limit equations
 \begin{equation}
 \partial_{t_m}^{}v(t^{}_m,x)=mv^{m-1}_{}(t^{}_m,x)v^{}_x(t^{}_m,x),\label{2:2}
 \end{equation}
i.e., the dispersionless KdV hierarchy. Solution of the initial problem for the $m$th equation
can be written in the parametric form as
 \begin{equation}
 x=s-mt^{}_mv_0^{m-1}(s),\qquad v(t^{}_m,x)=v(s),\label{2:3}
 \end{equation}
where $v(x)$ is initial data. This solution is known to describe overturn of the front, so the
initial problem for the Eqs.~(\ref{2:2}) has no global solution. On the other side,
Eq.~(\ref{51}) gives global solution of the quantum hierarchy~(\ref{44}). It is easy to see
that before the overturn of the front we get from~(\ref{51}) in the limit $\varepsilon\to 0$
(i.e., $\hbar\to 0$) that
 \begin{align}
 &v(t^{}_m,x)=\frac{1}{(2\pi)_{}^{2}i}\int dx'\int dy'\int dk\int dp\,\frac{e_{}^{iy'v(x-x')}-1}{y'}
 e_{}^{ikx'-ipy'+imt_m k p^{m-1}},
 \label{52}\\
\intertext{so that for the classical solution of~(\ref{22}) we get representation}
 &v(t^{}_m,x)=\int dp \,\bigl[\theta\bigl(v(x+mt_mp^{m-1}_{})-p\bigr)-\theta(-p)\bigr],
 \label{53}
\end{align}
that in this region coincides with~(\ref{2:3}) (here $\theta(p)$ denotes the step function).
Summarizing, it is natural to call the hierarchy introduced in the Sec.~3 \textbf{the quantum
dispersionless KdV hierarchy}. Dispersionless limits of integrable hierarchies attract now
essential attention in the literature, see~\cite{15,16}.

Our construction here is essentially based on the representation~\ref{21} valid for the
standard massless fermionic fields. Thanks to this relation we got description of the quantum
dispersionless KdV hierarchy. It is natural to hypothesize that anyonic
generalization~\cite{17} of the fermions leads to more generic integrable bozonic systems. 

\textbf{Acknowledgment.} The author thanks Isaac Newton Institute for Mathematical studies for
support of his participation in the program ``Integrable systems.'' He also thanks Prof.
P.~P.~Kulish for fruitful discussions.

\end{document}